\documentclass[twocolumn,aps,prl,showpacs]{revtex4}

\usepackage{graphicx}
\usepackage{dcolumn}
\usepackage{bm}

\begin{document}

\preprint{}

\title{Purely electronic THz polarization in dimer Mott insulators}

\author{Hiroki Gomi${}^{1,2}$}
\author{Takayuki Imai${}^{1}$}
\author{Akira Takahashi${}^{1,2}$}
\author{Masaki Aihara${}^{1}$}
\affiliation{${}^{1}$ Graduate School of Materials Science, Nara Institute of
Science and Technology, Ikoma, 630-0192, Japan}
\affiliation{${}^{2}$ CREST, Japan Science and Technology Agency,
Chiyoda-ku, Tokyo 102-0075, Japan}

\date{\today}

\begin{abstract}
We theoretically discover purely electronic polarization modes in THz
frequency region in dimer Mott insulators $\kappa$-(BEDT-TTF)$_2$X.
The unusual low-frequency modes arise from the coupling between
the oscillation of intradimer electric dipole moments and that of
alternating interdimer bond orders.
These collective motions play an important
role in the dynamical dielectric properties of the dimer Mott insulators.
Near the phase boundary of the dimer Mott transition, the ferroelectric
ground state is realized by introducing electron-lattice
coupling.
\end{abstract}


\maketitle
Low-dimensional strongly correlated electron systems with a
quarter-filled band have recently attracted much attention.
BEDT-TTF salts are
typical two-dimensional systems with
anisotropic triangular lattices, and the strong correlation and
frustration act cooperatively to generate exotic phases.
Among the BEDT-TTF salts, $\kappa$-(BEDT-TTF)$_2$X (X: a counter anion)
exhibit the Mott insulator phase \cite{K1,K2,K4,K5,K6},
exotic superconducting phase \cite{SC}, spin liquid phase
\cite{SL,HA1,HA2}, and so on. The BEDT-TTF molecular lattice is
distorted to form dimers in $\kappa$-(BEDT-TTF)$_2$X.
Hybridized two
sites can be effectively treated as a single site, and the valence band
is regarded as a half-filled one. As a result, $\kappa$-(BEDT-TTF)$_2$X
exhibits the Mott insulator phase if the effective on-site Coulomb
energy is large enough. The state is called the dimer Mott insulator.

Recently quite interesting charge excitation phenomena have been observed
in the dimer Mott insulators: photoinduced phase transition to a metallic
state \cite{PIPT}, and anomaly in dielectric properties \cite{DA}.
In particular,
the latter suggests that low-energy charge excitation exists in
the dimer Mott insulators in contrast to the case of the Mott
insulators.

In this letter, we show for the first time that the dimer Mott insulator
has THz electric polarization, which purely arises from
the electron dynamics.
The ferroelectricity which arises from the electron dynamics has been
observed in some materials, and these ferroelectrics have unconventional
origins such as charge order generation \cite{F1,F2,F3,F4,F5}.
We propose another new
origin of ferroelectricity in this paper.

We adopt the 1/4-filled extended Hubbard Hamiltonian for holes on the
two-dimensional anisotropic triangular lattice, which includes the degrees
of freedom in a dimer.
It is given by
\begin{eqnarray} \label{eq:H}
H=\sum_{<n, m>} (\beta_{n, m}
{\hat p}_{n,m} +  V_{n,m} n_{n} n_{m})
 + U \sum_{n} n_{n, \uparrow} n_{n, \downarrow},
\end{eqnarray}
where $<n, m>$ denotes the neighbor site pairs,
$c_{n, \sigma}$ ($c_{n, \sigma}^{\dag}$) is the annihilation
(creation) operator for a hole of spin $\sigma$ at the site $n$,
$\beta_{n, m}$ ($V_{n, m}$) is the transfer integral
(the Coulomb interaction energy) between the sites
$n$ and $m$,
${\hat p}_{n,m}=\sum_{\sigma} (c_{m, \sigma}^{\dag}c_{n, \sigma}
+ c_{n, \sigma}^{\dag}c_{m, \sigma})$,
$U$ is the on-site Coulomb interaction energy,
$n_{n, \sigma}=c_{n, \sigma}^{\dagger} c_{n, \sigma}$,
and $n_{n}=\sum_{\sigma} n_{n, \sigma}$.
The $4 \times 4$ cluster of the system size $N=16$
with the periodic boundary condition
is used.

We calculate the quantum state excited by light pulse
expressed by the vector potential
$\bm{A}(t)=\bm{e} A \exp(-(t/D)^2) \cos(\omega_{\rm p} t)$
of a Gaussian profile.
We consider the one-photon excited states in this paper.
To the first order in $A$,
the solution $|\psi (t) \rangle$ of the Schr\"odinger equation
subject to the pulse is given by
$|\psi (t) \rangle = |\psi_{\rm 0} \rangle +
|\psi^{(1)} (t) \rangle$.
The ground state $|\psi_{\rm 0} \rangle$ and the first order solution
$|\psi^{(1)} (t) \rangle$ are numerically calculated.

We consider the first-order changes of
the hole density $\rho ^{(1)}_{n} (t) = (\langle \psi_{\rm 0} | n_{n}
|\psi^{(1)} (t) \rangle + c.c.)/A$ and the bond order
$p ^{(1)}_{n,m} (t) =  (\langle \psi _{\rm 0} | {\hat p}_{n,m}
|\psi^{(1)} (t) \rangle + c.c.)/A$
induced by the pulse excitation.
Since the absolute value of the bond order shows the strength of the bond,
we define the difference in the absolute value of the bond orders between
$|\psi (t) \rangle$ and $|\psi_{\rm 0} \rangle$ as
$|p| ^{(1)}_{n,m} (t) =
|{\bar p}_{n,m} + p ^{(1)}_{n,m} (t)|
-|{\bar p}_{n,m}|,$
where ${\bar p}_{n,m}$ is the bond order for the ground state.
The bond is stronger in $|\psi (t) \rangle$ than in $|\psi_{\rm 0} \rangle$
for $|p| ^{(1)}_{n,m} > 0$.

\begin{figure}
\includegraphics{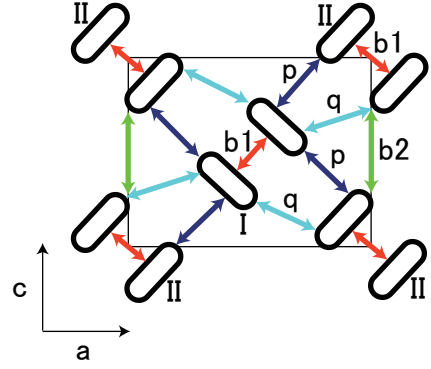}
\caption{(Color online)
Anisotropic triangular lattices for $\kappa$-(BEDT-TTF)$_2$X.}
\label{fig:ATL}
\end{figure}

As shown in Fig. \ref{fig:ATL}, there are four nonequivalent bonds,
and they are labeled by b1, b2, p, and q according to Mori {\it et al.}
\cite{Lattice}.
The b1 bond is much stronger than the other ones, and the two BEDT-TTF
molecules connected by the b1 bond form a dimer.
There are two dimers in the unit cell, and they are labeled by I and II
as shown in this figure. The transfer integral
$\beta_{\rm b1}$ and $V_{\rm b1}$ denote
$\beta_{n, m}$ and $V_{n, m}$ for the b1 bond, respectively,
and the others are denoted in the same way.
The transfer integrals for b1, p, and q bonds are fixed to the following
values for $\kappa$-(BEDT-TTF)$_2$Cu[N(CN)$_2$]Br deduced from
the extended H\"uckel calculation \cite{Lattice}:
$\beta_{\rm b1}=0.265$ eV, $\beta_{\rm p}=0.109$ eV, and
$\beta_{\rm q}=-0.038$ eV.
As for the Coulomb parameters, we use the following values:
$U=0.7$ eV \cite{KF1,KF3},
$V_{\rm b1}=0.45$ eV, $V_{\rm b2}=V_{\rm p}=V_{\rm q}=0.25$ eV.
With these fixed parameters, we show the $\beta_{\rm b2}$ dependence
of various physical quantities in this paper
because the spin frustration
can be controlled by changing $\beta_{\rm b2}$.

We first show the physical properties of the ground state.
The spin correlation function discontinuously
changes at $\beta_{\rm b2} =$ 0.073 eV.
For $\beta_{\rm b2} \leq$ 0.073 eV, the spin correlation is always
negative between the dimers I and II, showing that
the ground state is an antiferromagnetic (AFM) state.
For $\beta_{\rm b2} \geq$ 0.074 eV, the spin correlations are much
smaller than those for the AFM ground state, showing that
the ground state is a paramagnetic (PM) state.
Decreasing $\beta_{\rm b1}$, the ground state discontinuously
changes from the AFM state to the PM state. The PM ground state changes
continuously as $\beta_{\rm b1}$
is decreased even to the value of $\beta_{\rm b2}$ or $\beta_{\rm p}$.
This shows that the PM state is stable without dimerization, and therefore
it is not a dimer Mott insulator of a spin liquid state.
We also calculate the probability $W_{\rm s}$ that
a dimer is singly occupied. The probability $W_{\rm s}$
also discontinuously changes at the phase boundary, and
$W_{\rm s} \geq 0.85$ holds all through the AF ground state region.
This indicates that the AFM state is a dimer Mott insulator.
On the contrary, $W_{\rm s}$ decreases continuously to about 0.5
as $\beta_{\rm b1}$ is decreased.
This result strongly suggests that the PM state is metallic.
In the realistic parameter region, the ground state is either
the AFM dimer Mott insulator or the PM metal, and
the phase transition between them occurs by changing
various parameters. This result is consistent with previous ones
\cite{KF1,KF3}.
The physical properties of these AFM and PM states
are basically unchanged in the realistic parameter range,
and the essentially same results are obtained.

\begin{figure}
\includegraphics{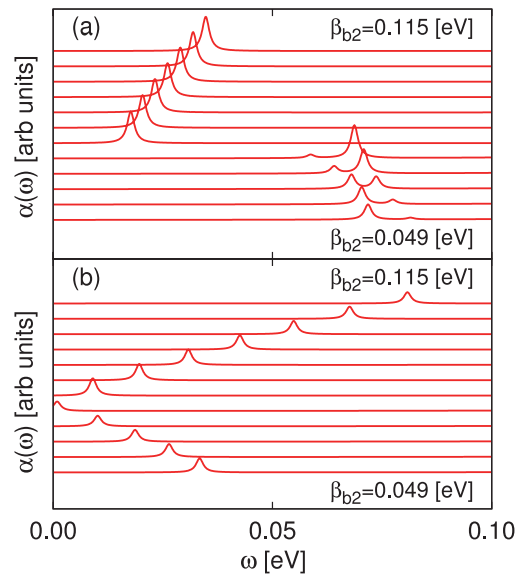}
\caption{(Color online)
Light absorption spectrum $\alpha (\omega)$ when the light is polarized
(a) to the $a$-direction and (b) to the $c$-direction for various values
of $\beta_{\rm{b2}}$.The interval of $\beta_{\rm{b2}}$ is 0.006 eV.}
\label{fig:Abs}
\end{figure}

In Fig. \ref{fig:Abs}, we show the $\beta_{\rm{b2}}$ dependence of
light absorption spectrum $\alpha (\omega)$
in the energy region of THz light.
The dominant peaks in $\alpha (\omega)$ are in the region
$\omega \gtrsim 0.2$ eV. In addition to these large peaks, there are
much smaller peaks in the THz region both
in the cases of light excitation from the AFM and PM ground states.
It should be emphasized that
such low energy peaks are not observed in the simple Mott and
charge ordered insulators.

We next show the time dependence of physical quantities induced by
a THz light pulse.
In all the following cases, the center frequency $\omega_{\rm p}$
of the pulse is chosen to be resonant to the largest peak
in the energy region shown in Fig. \ref{fig:Abs}. The used duration
$D= 100$ eV${}^{-1}$ is large enough that the contributions
from the large peaks in the higher energy region are negligible.

\begin{figure}
\includegraphics{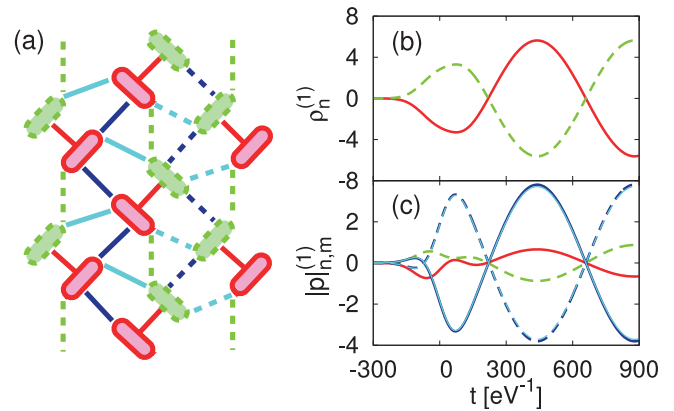}
\caption{(Color online)
The charge and bond order induced by the THz light polarized
along the $c$-direction for $\beta_{\rm b2}=$0.068 eV.
(a) Nonequivalent bonds and sites are distinguished by colors and
kinds (solid or dashed) of lines.
(b) $\rho ^{(1)} _{n} (t)$ and (c) $|p| ^{(1)}_{n,m}(t)$
are shown by the lines of the colors and kinds that specify the site $n$
and the bond $n,m$, respectively, in (a).
}
\label{fig:rho1c}
\end{figure}

We first show the results in the case of the excitation from
the AFM ground state. In Fig. \ref{fig:rho1c}, we show
the time dependence of charge and bond order
induced by the THz light when the light is
polarized to the $c$-axis at $\beta_{\rm{b2}}=0.068$ eV.
The b1, b2, p and q bonds are shown by the red, the green, the dark blue,
and the light blue lines, respectively, in Fig. \ref{fig:rho1c} (a).
Moreover, there are two nonequivalent
p (q) bonds with different bond orders in the photoexcited state
$|\psi (t) \rangle$. These two nonequivalent bonds are distinguished
by the solid and the dashed lines.
Two sites forming a dimer also become nonequivalent, and they are
distinguished by the solid red and the dashed green ellipsoids.
We show $\rho ^{(1)} _{n} (t)$ and $|p| ^{(1)}_{n,m}(t)$ in
Figs. \ref{fig:rho1c} (b) and (c), respectively, for the sites and
the bonds shown in Fig. \ref{fig:rho1c} (a).
As seen from these figures, $\rho ^{(1)} _{n} (t)$ and
$|p| ^{(1)}_{n,m}(t)$ oscillate coherently.
The quantities shown by the solid lines and the dashed lines oscillate
out of phase.

\begin{figure}
\includegraphics{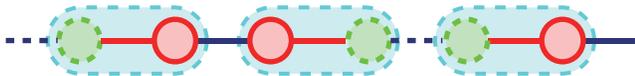}
\caption{(Color online)
The schematic representation of bond order alternation between dimers.
The ellipsoids represent the dimers, and
the solid (dashed) lines show the stronger (weaker) bonds
connecting the dimers. The solid (dashed) circles represent
the charge-rich (charge-poor) sites.}
\label{fig:SDD}
\end{figure}

Before analyzing the present results, we introduce
bond order alternation between dimers in much simpler 1D
quarter-filled electron-lattice coupled systems.
The dimer Mott insulator with alternating bond
lengths and bond orders between the dimers is schematically shown
in Fig. \ref{fig:SDD}. 
The site connected by the stronger bond becomes charge-rich because
electrons are more stabilized on the stronger bond. The state
becomes the ground state with the deformed lattice in a parameter range
\cite{SD1,SD2,SD3}.

The collective motion induced by the THz light pulse can be regarded
as the bond order alternation between the dimers.
As seen from Fig. \ref{fig:rho1c} (c),
the amplitudes of the induced bond order oscillation for the p and q bonds
are much larger than those for the other bonds, and therefore we
focus on the p and q bonds.
As seen from Fig. \ref{fig:rho1c} (a),
the interdimer bond orders alternate
along the $a$-direction, and the amplitude of the alternation
oscillates with time.
Since charges are attracted by stronger bonds, when the bonds shown
by the solid (dashed) lines become stronger, the sites shown
by the solid red (dashed green) ellipsoids become charge-rich as seen
from Fig. \ref{fig:rho1c} (b). In this way,
the bond order and the charge density oscillations are strongly coupled in
the collective mode, and the energy of the charge
oscillation is significantly reduced by the coupling.
The excitation induces the polarization along the $c$-direction
due to the electric dipole moment on each dimer.

It should be emphasized that the lattice deformation
is not considered here. In the previous works,
interdimer bond order alternation occurs as
a result of electron-lattice coupling.
On the other hand,
in the present case, the induced collective motion of
the bond order alternation coupled with
polarization has purely electronic origin.
We have found this purely electronic low-energy excitation also
in the one-dimensional dimer Mott insulators without
electron-lattice coupling.

\begin{figure}
\includegraphics{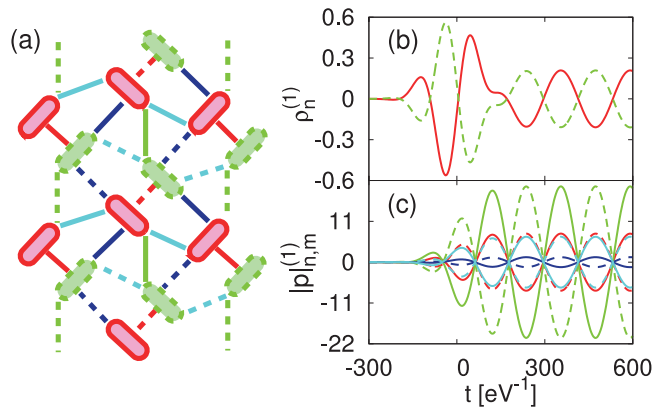}
\caption{(Color online)
The charge and bond order dynamics induced by the THz light polarized
along the $a$-direction for $\beta_{\rm b2}=$0.098 eV.
(a), (b), and (c) The same as Fig. \ref{fig:rho1c}.
}
\label{fig:rho1aP}
\end{figure}

When the light is polarized to the $a$-axis, the collective motion of
the interdimer bond order alternation is induced by
the THz light pulse.
The collective motion is just like the case of the excitation from
the PM ground state schematically shown in Fig. \ref{fig:rho1aP} (a).
The bond orders for
the p and q bonds alternate along the $c$-direction.
The bond orders and the electric dipole moments
oscillate coherently in the mode, and the electric polarization along
the $a$-direction is induced by the excitation.

We next show the results in the case of the excitation from
the PM ground state.
When the light is polarized to the $c$-axis, 
the collective motion shown in Fig. \ref{fig:rho1c} (a) is induced 
just like the case of the AFM ground state.

In Fig. \ref{fig:rho1aP}, we show the result when the light is
polarized to the $a$-axis at $\beta_{\rm{b2}}=0.098$ eV. As seen from
this figure, the amplitude of the induced oscillation for the b2 bonds
is the largest between the bonds in this case, and it is much larger
than the largest ones in the other cases.
However, the induced
dipole moment is much smaller in this case compared with the other cases.
This can be understood as follows. As seen from Fig. \ref{fig:rho1aP} (a),
a dimer is connected by two solid b2 bonds or two dashed b2 bonds.
The dominant motion of the b2 bonds does not accompany
the bond order alternation between the dimers, and the coupling with
the charge motion is very weak.

As a result of electronic degrees of freedom in a dimer, weak charge
disproportion is coupled with the interdimer bond order alternation.
Therefore,
there exist no such low-energy charge excitations in the Mott insulators.
Furthermore, these collective modes occur
irrespective of the AF spin order and the Mott insulating phase.
Spin frustration also does not play an important role for these modes.
The formation
of the dimer and strong correlation effects are essential for these modes.

The ferroelectric phase is not obtained within the present model.
However, the ground state becomes ferroelectric by introducing
electron-lattice coupling. For example, if the dimers denoted by I (II)
in Fig. \ref{fig:ATL} shift to left (right), then
the transfer integrals for the p (q) bonds become $\beta_{\rm p}+s\Delta$
($\beta_{\rm q}-sr\Delta$), where $\Delta>0$
shows the magnitude of bond length alternation, $s=1$ ($-1$) for the bonds
shown by the solid (dashed) lines, and $r>0$ accounts for the difference
in the electron-lattice coupling. Since
$\beta_{\rm p}>0$ and $\beta_{\rm q}<0$, the sign for the coupling term
is opposite.
We consider the state
$|\psi \rangle=\sqrt{1-C_1^2}|\psi_0 \rangle+C_1 |\psi_1 \rangle$,
where $C_1$ is a real constant, 
$|\psi_1 \rangle = \langle \psi^{(1)} (\tau) | \psi^{(1)} (\tau) \rangle^{-1/2}
|\psi^{(1)} (\tau) \rangle$ when the light is polarized to the $c$-direction,
and $|p| ^{(1)}_{n,m}(t)$
for the bonds shown by the solid lines become maximum at $t=\tau$.
The energy eigenstate $|\psi_1 \rangle$ is responsible to the absorption
peak in the THz frequency region.
The bond orders for
$|\psi \rangle$ for the x bonds are
${\bar p}_{\rm x}(1-C_1^2)+sp_{\rm x}^{\prime}C_1\sqrt{1-C_1^2}
+(1/2)p_{\rm x}^{\prime\prime}C_1^2$,
where x=p or q,
${\bar p}_{\rm x}=\langle \psi _{\rm 0} |{\hat p}_{n,m}|\psi_0 \rangle$,
$p_{\rm x}^{\prime}=
\langle \psi _{\rm 0} |{\hat p}_{n,m}|\psi_1 \rangle + c.c.$ and
$p_{\rm x}^{\prime\prime}=2
\langle \psi _{\rm 1} |{\hat p}_{n,m}|\psi_1 \rangle$.
The energy expectation value for $|\psi \rangle$
with the distorted lattice minus the ground state energy $E_0$ without
lattice distortion is given by
$\Delta E (C_1,\Delta)=\alpha C_1^2+\beta C_1 \Delta + \gamma \Delta^2$,
to the second order in $C_1$ and $\Delta$.
Here $\alpha=E_1 - E_0$, $E_1$ is the energy eigenvalue for $|\psi_1 \rangle$,
$\beta=N_{\rm p}(p_{\rm p}^{\prime}-rp_{\rm q}^{\prime})$, $N_{\rm p}$ is
the number of p or q bonds,
$\gamma=N_{\rm p}(1/2)(K_{\rm p}+K_{\rm q} r^2)$, and
$K_{\rm x}$ is the renormalized spring constant for x
bond. $\Delta E$ is minimized
when $C_1=-\beta/(2\alpha)\Delta$
and then $\Delta E=(\gamma-\beta^2/(4\alpha))\Delta^2$. 
Therefore, when
$\alpha<4\beta^2/\gamma$, lattice distortion occurs and $|\psi_1 \rangle$ is
hybridized to $|\psi _{\rm 0} \rangle$. 
As a result,
the ground state have a finite electric polarization.
In the present case, $|\beta|$ is very large because
the contributions from all the p and q bonds are negative as a result
of the collective nature of the excitation.
It is estimated that $K_{\rm x} \sim 500$ eV${}^{-1}$
for $\alpha$-(BEDT-TTF)$_2$X \cite{EL1}.
Using this value and setting $r=1$, we obtain $4\beta^2/\gamma=0.03$ eV.
Since $\alpha$ is close to
zero near the phase boundary, the ferroelectric phase is realized
near the phase boundary, which is confirmed
by a numerical calculation.

With using the effective Hamiltonian where the interdimer part is treated
perturbatively, the ferroelectric ground state is obtained without
lattice deformation \cite{FK}, and this is not consistent with our result.
We consider that this is because the interdimer part cannot be treated
perturbatively.
The effective Hamiltonian for the strong correlation case is also
proposed \cite{EH}.

Based on the present results, dielectric anomaly observed in
$\kappa$-(BEDT-TTF)$_2$Cu$_2$(CN)$_3$ may be explained in the
following way. As the temperature $T$ is decreased, the system becomes
close to the phase boundary. As a result, the domains of the ferroelectric
phase are generated for $T<60$K, which results in the broad peak with
relaxor like relaxation in dielectric constant.
Around $T=6$K, the ground state becomes ferroelectric,
and this is the origin of the anomaly of various physical quantities
\cite{SL,HA1,HA2,LA1,LA2}.
In particular, the present result is consistent with the
unexpected experimental result that anomalous lattice response is
observed only along the $a$-direction \cite{LA1,LA2}.
Similar scenarios are proposed also in Refs. \cite{FK,MN}.

Motional spectrum narrowing of an intermolecular phonon mode
is observed in $\kappa$-(BEDT-TTF)$_2$Cu$_2$(CN)$_3$ around
$T=30$K \cite{MN}, and this can be attributed to the collective modes
presented in this paper. These modes are strongly coupled with
the phonon mode of interdimer bond length alternation.
This is consistent with the fact that the narrowing occurs only
in one specific phonon mode. Narrowing does not occur for $T<5$K.
This is because the ground state becomes ferroelectric and
these modes disappear in the temperature region. Furthermore,
the correlation time of the fluctuation mode roughly agrees with
the energies of these collective modes apart from the phase boundary.

\begin{acknowledgments}
This research was supported by CREST, JST.
\end{acknowledgments}


\begin{thebibliography}{99}
\bibitem{K1} T. Sasaki, I. Ito, N. Yoneyama, N. Kobayashi, N. Hanasaki,
H. Tajima, T. Ito, and Y. Iwasa, 
 Phys. Rev. B {\bf 69}, 064508 (2004).
\bibitem{K2} F. Kagawa, K. Miyagawa, and K. Kanoda, 
 Nature, {\bf 436}, 534 (2005).
\bibitem{K4}  D. Faltermeier, J. Barz, M. Dumm, M. Dressel,N. Drichko,
B. Petrov, V. Semkin, R. Vlasova, C. Me\'{z}i\`{e}re, and P. Batail,
Phys. Rev. B {\bf 76}, 165113 (2007).
\bibitem{K5} T. Sasaki, N. Yoneyama, Y. Nakamura, N. Kobayashi,
Y. Ikemoto, T. Moriwaki, and H. Kimura,
Phys. Rev. Lett. {\bf 101}, 206403 (2008).
\bibitem{K6} M. Dumm, D. Faltermeier, N. Drichko, M. Dressel,
C. M\'{e}zi\`{e}re, and P. Batail,
Phys. Rev. B {\bf 79}, 195106 (2009).
\bibitem{SC} M. S. Nam, A. Ardavan, S. J. Blundell, and A. Schlueter,
Nature {\bf 449}, 584 (2007).
\bibitem{SL} Y. Shimizu, K. Miyagawa, K. Kanoda, M. Maesato, and G. Saito, 
 Phys. Rev. Lett. {\bf 91} (2003) 107001
\bibitem{HA1} S. Yamashita, Y. Nakazawa, M. Oguni, Y. Oshima, H. Nojiri,
Y. Shimizu, K. Miyagawa and K. Kanoda, Nature Physics {\bf 4}, 459 (2008)
\bibitem{HA2} M. Yamashita, N. Nakata, Y. Kasahara, T. Sasaki,
N. Yoneyama, N. Kobayashi, S. Fujimoto, T. Shibauchi, and Y. Matsuda, 
Nature Physics {\bf 5}, 44-47 (2009).
\bibitem{PIPT} Y. Kawakami, S. Iwai, T. Fukatsu, M. Miura,
N. Yoneyama, T. Sasaki, and N. Kobayashi,
Phys. Rev. Lett. {\bf 103}, 066403 (2009).
\bibitem{DA} M. Abel-Jawad, I. Terasaki, T. Sasaki, N. Yoneyama,
N. Kobayashi, Y. Uesu, and C. Hotta,
unpublished.
\bibitem{F1} P. Monceau, F. Ya. Nad, S. Brazovskii,
Phys. Rev. Lett. {\bf 86}, 4080 (2001).
\bibitem{F2} N. Ikeda, H. Ohsumi, K. Ohwada, K. Ishii, T. Inami, K. Kakurai,
Y. Murakami, K. Yoshii, S. Mori, Y. Horibe and H. Kito,
Nature, {\bf 436}, 1136 (2005).
\bibitem{F3} A. Nagano, M. Naka, J. Nasu, and S. Ishihara,
Phys. Rev. Lett. {\bf 99}, 217202 (2007).
\bibitem{F4} M. Naka, A. Nagano, and S. Ishihara, 
Phys. Rev. B {\bf 77}, 224441 (2008).
\bibitem{F5} K. Yamamoto, S. Iwai, S. Boyko, A. Kashiwazaki, F. Hiramatsu,
C. Okabe, N. Nishi, and K. Yakushi, 
J. Phys. Soc. Jpn. {\bf 77}, 074709 (2008).
\bibitem{Lattice} T. Mori, H. Mori, and S. Tanaka, 
Bull. Chem. Soc. Jpn. {\bf 72}, 179 (1999).
\bibitem{KF1} H. Kino and H. Fukuyama,
J. Phys. Soc. Jpn. {\bf 64}, 2726 (1995).
\bibitem{KF3} H. Kino and H. Fukuyama,
J. Phys. Soc. Jpn. {\bf 65}, 2158 (1996).
\bibitem{SD1} R. T. Clay, S. Mazumdar, and D. K. Campbell,
Phys. Rev. B {\bf 67}, 115121 (2003).
\bibitem{SD2} M. Kuwabara, H. Seo and M. Ogata,
J. Phys. Soc. Jpn. {\bf 72}, 225 (2003).
\bibitem{SD3} Y. Otsuka, H. Seo, Y. Motome, and T. Kato, J. Phys. Soc.
Jpn. {\bf 77}, 113705 (2008).
\bibitem{EL1} Y. Tanaka and K. Yonemitsu,
J. Phys. Soc. Jpn. {\bf 77}, 094712 (2008).
\bibitem{FK} M. Naka, and S. Ishihara, arXiv.1002.4971.
\bibitem{EH} C. Hotta, arXiv:0912.3674v1.
\bibitem{LA1} M. de Souza, A. Br{\"{u}}hl, Ch. Strack, B. Wolf,
D. Schweitzer, and M. Lang, 
 Phys. Rev. Lett. {\bf 99}, 037003 (2007).
\bibitem{LA2} R. S. Manna, M. de Souza, A. Br\"{u}hl, J. A. Schlueter, and
M. Lang: arXiv:0909.0718.
\bibitem{MN} K. Itoh, H. Nakaya, Y. Kawakami, T. Fukatsu, H. Itoh,
S. Iwai1, T. Sasaki, and S. Saito, unpublished.
\end{thebibliography}
\end{document}